\documentclass[12pt,twoside]{article}
\usepackage{epsfig}
\begin{document}\hbadness=10000\thispagestyle{empty}
\pagestyle{myheadings}
\title{Quantum Mechanics and Discrete Time \\ 
from ``Timeless'' Classical Dynamics\footnote{Lecture given at DICE 2002. To be published in: 
{\it Decoherence and Entropy in Complex Systems}, \protect\newline Lecture Notes in Physics (Springer-Verlag, Berlin 2003).}}
\author{$\ $\\
{Hans-Thomas Elze} 
\\ $\ $\\ 
Instituto de F\'{\i}sica, Universidade Federal do Rio de Janeiro \\ 
C.P. 68.528, 21941-972 Rio de Janeiro, RJ, Brazil \\ 
E-mail: thomas@if.ufrj.br
}
\vskip 0.5cm
\date{July 2003}
\maketitle
\vspace{-8.5cm}
\vspace*{8.0cm}
\begin{abstract}{\noindent
We study classical Hamiltonian systems in which the intrinsic 
proper time evolution parameter is related through a probability 
distribution to the physical time, which is assumed to be discrete. 

This is motivated by the ``timeless'' reparametrization invariant 
model of a relativistic particle with two compactified extradimensions. 
In this example, discrete physical time is constructed based on 
quasi-local observables. 

Generally, employing the path-integral formulation of classical 
mechanics developed by Gozzi et al., we show that these deterministic 
classical systems can be naturally described as unitary quantum mechanical 
models. The emergent quantum Hamiltonian is derived from the underlying 
classical one. It is closely related to the Liouville operator. We 
demonstrate in several examples the necessity of regularization, in order 
to arrive at quantum models with bounded spectrum and stable groundstate.
}\end{abstract}
\newpage
\section{Introduction}
Recently we have shown that for a particle with 
time-reparametrization invariant dynamics, be it relativistic or nonrelativistic, 
one can define quasi-local observables 
which characterize the evolution in a gauge invariant way \cite{ES02,E03}. 
  
We insist on quasi-local measurements in describing 
the evolution, which respect reparametrization 
invariance of the system. Then, as we have argued, the physical 
time necessarily becomes discrete, 
its construction being based on a Poincar\'e section which reflects ergodic dynamics,   
by assumption. 
Most interestingly, due to inaccessability of globally complete information on trajectories, the  
evolution of remaining degrees of freedom appears as in a quantum mechanical model 
when described in relation to the discrete physical time.     
  
While we pointed out in the explicitly ergodic examples of references \cite{ES02,E03} that such 
emergent discrete time leads to what may, for obvious reasons, be called ``stroboscopic'' quantization,  
we report here how this occurs quite generally in classical Hamiltonian systems, 
if time is discrete and related to the proper time of the equations of motion 
in a statistical way \cite{E036}. In our concluding section, we will briefly comment about  
extensions, where the prescribed probabilistic mapping of physical onto proper time 
shall be abandoned in favour of a selfconsistent treatment. A closed system has   
to include its own ``clock'', if it is not entirely static, reflecting the experience  
of an observer in 
the Universe.

Previous related work on the ``problem of time'' has 
always assumed that global 
features of the trajectory of the system are accessible to the observer.  
This makes it possible, in principle, to 
express the evolution of an arbitrarily selected degree of freedom relationally in terms of others 
\cite{MRT99,Montesinos00}. Thereby the Hamiltonian and possibly additional constraints have been eliminated 
in favour of Rovelli's ``evolving constants of motion'' \cite{Rovelli90}. For a recent development aiming at solving  
the constraints after discretization see \cite{Pullin}. 
While appealing by its conceptual clarity, incorporating nonlocal observations seems unrealistic to us 
in any case. 

In distinction, we point out that the emergent discrete time in our approach naturally 
leads to the ``stroboscopic'' quantization of the system \cite{ES02,E03,E036}. Quantum theory thus appears to originate from ``timeless'' classical dynamics, due to the lack of globally complete information \cite{E036}.

Another approach to deterministically induced quantization is proposed in \cite{Wetterich02}, 
where the consequences of incomplete statistics are analyzed, leading 
towards Euclidean quantum field theory under very general assumptions. 
Various other arguments considering quantization 
as an emergent property of classical systems have recently been proposed, for example, 
in references \cite{tHooft01,Vitiello,Mueller,Smolin,Adler}, concerning quantum gravity and dissipation 
at a fundamental level, ``chaotic quantization'', and matrix models, respectively.  

Our approach tries to 
illuminate from a different angle how to arrive at quantum 
models which describe dynamically evolving systems. In particular, 
we believe that there may be an intimate connection with how the ``problem of time'' 
is resolved for a local observer, namely by counting suitably defined and locally  
measurable incidents.    

We remark that the possibility of a fundamentally discrete time (and possibly other discrete 
coordinates) has been explored before, ranging from an early realization of Lorentz 
symmetry in such a case \cite{Snyder} to detailed explorations of its consequences 
and consistency in classical mechanics, quantum field theory, and general relativity 
\cite{TDL83,JN97,Pullinetal}. 
However, no detailed models giving rise to such discreteness have been proposed.  
Quantization, then, is always performed in an additional step, as usual.  

In particular, the work by Gambini, Pullin et al. aims at a consistent canonical quantization of 
gravity via discretization \cite{Pullin,Pullinetal}. Discretization of time is performed 
in a {\it static} fashion, i.e. independently of the evolution. As shown there, the major advance 
lies in the possibility to satisfy the constraints, in principle, by suitably choosing the 
Lagrange multipliers. However, the extraneous discretization is reflected by the persistence of 
a discrete variable $n$ after quantization, which apparently has no physical meaning.  

Presently, we shall see a vague resemblance to the persistence of proper time $\tau$ 
in our approach, as long as the discrete physical time is related through a {\it given} probability 
distribution to proper time. 
Our formalism, however, is set up in such a way that the ``clock'' degrees of freedom can 
be treated dynamically as part of the system. 
This should help to pinpoint the role of proper time in the resulting  
model where, in our case, {\it quantization emerges} instead of being imposed on the system.     

In the following section, we recall the model from \cite{E03}, in order to motivate 
the emergence of discrete time from quasi-local observations. This is the starting point 
of our heuristic derivation of a quantum mechanical picture of what appear fundamentally 
classical systems.    
   
To put our approach into perspective, we remark that there is clearly no need to 
follow such construction leading to a discrete physical time in ordinary mechanical systems or 
field theories, where time is an external classical parameter, commonly called ``t''. 
However, assuming for the time being that truly fundamental theories 
will turn out to be diffeomorphism invariant, adding further the requirement of 
the observables to be quasi-local (modulo a fundamental length scale), when describing the evolution, then 
such an approach seems natural, which may lead to quantum mechanics as an 
emergent description or ``effective theory'' on the way.     

\section{Discrete Time of a Relativistic Particle \protect\newline with Extradimensions}
We consider the (5+1)-dimensional model of a ``timeless'' relativistic particle 
(rest mass $m$) with the action:  
\begin{equation}\label{S}
S=\int\mbox{d}s\;L
\;\;, \end{equation}
where the Lagrangian is defined by: 
\begin{equation}\label{Lp}
L\equiv -\frac{1}{2}(\lambda^{-1}\dot x_\mu\dot x^\mu +\lambda m^2)
\;\;. \end{equation}
Here $\lambda$ stands for an arbitrary ``lapse'' function of the evolution parameter $s$, 
$\dot x^\mu\equiv\mbox{d}x^\mu /\mbox{d}s\,(\mu =0,1,\dots,5)$, and the metric is 
$g_{\mu\nu}\equiv\mbox{diag}(1,-1,\dots,-1)$. Units are such that $\hbar =c=1$.    

With this form of the Lagrangian, instead of the frequently encountered 
$L\mbox{d}s\propto (g_{\mu\nu}\mbox{d}x^\mu\mbox{d}x^\nu )^{1/2}$ 
which emphasizes the geometric (path length) character of the action, the presence of a constraint is immediately 
obvious, since there is no $s$-derivative of $\lambda$. 

Two spatial coordinates, $x^{4,5}$ in (\ref{Lp}), are toroidally compactified: 
\begin{eqnarray}\label{compact1}
x^{4,5}&\equiv&2\pi R[\phi^{4,5}] 
\;\;, \\ [1ex] \label{compact2}
[\phi ]&\equiv&\phi -n\;\;,\;\;\;\phi\in [n,n+1[
\;\;, \end{eqnarray} 
for any integer $n$, i.e. the angular variables are periodically continued; 
henceforth we set $R=1$, for convenience.
Alternatively, we can normalize the angular variables to the square $[0,1[\times [0,1[$, 
of which the opposite boundaries are identified, thus describing the surface of a torus with main radii equal to one.  

While full Poincar\'e invariance is broken, as in other currently investigated models 
with compactified higher dimensions, the usual one remains in fourdimensional Minkowski space together 
with discrete rotational invariance in the presently two extradimensions; also 
translational symmetry persists. Furthermore,  
the internal motion on the torus is ergodic with an uniform asymptotic density 
for almost all initial conditions, in particular if the ratio of the corresponding initial momenta 
is an irrational number. 
    
Setting the variations of the action to zero, we obtain:  
\begin{eqnarray}\label{lapse}
\frac{\delta S}{\delta\lambda}&=&
\frac{1}{2}(\lambda^{-2}\dot x_\mu\dot x^\mu -m^2)=0
\;\;, \\ [1ex] \label{x}
\frac{\delta S}{\delta x_\mu}&=&
\frac{\mbox{d}}{\mbox{d}s}(\lambda^{-1}\dot x^\mu)=0
\;\;. \end{eqnarray}  
In terms of the canonical momenta, 
\begin{equation}\label{momentum} 
p_\mu\equiv\frac{\partial L}{\partial\dot x^\mu}=-\lambda^{-1}\dot x_\mu 
\;\;, \end{equation} 
the equations of motion (\ref{x}) become simply $\dot p^\mu =0$, while (\ref{lapse}) 
turns into the mass-shell constraint $p^2-m^2=0$. 

The equations of motion are solved by: 
\begin{equation}\label{xsolution}  
x^\mu (s)=x_i^\mu -p^\mu\int_0^s\mbox{d}s'\lambda (s')\equiv x_i^\mu +p^\mu\tau(s)
\;\;, \end{equation} 
where the conserved (initial) momentum $p^\mu$ is constrained to be on-shell and 
$x_i$ denotes the initial position. Here we 
also defined the fictitious proper time (function) $\tau$, which allows us to 
formally eliminate the lapse function $\lambda$ from (\ref{lapse})-(\ref{x}), 
using $x^\mu (\tau )\equiv x^\mu (s)$ and  
$\dot x^\mu (s)=-\lambda (s)\partial_\tau x^\mu (\tau )$.    
  
In order to arrive at a physical space-time 
description of the motion, the proper time  
needs to be determined in terms of observables. In the simplest case, the result should be given by 
functions $x^{\mu\neq 0}(x^0)$, provided there is a physical clock measuring  
$x^0=x_i^0+p^0\tau$.  
  
Similarly as in the nonrelativistic example studied in \cite{ES02}, 
the lapse function introduces a gauge degree 
of freedom into the dynamics, which is related to the reparametrization of the evolution parameter $s$. 
In fact, the action, (\ref{S})-(\ref{Lp}), is invariant under the set of gauge 
transformations: 
\begin{equation}\label{gauge} 
s\equiv f(s')\;\;,\;\;\; x^\mu (s)\equiv x'^\mu (s')\;\;,\;\;\; \lambda (s)\frac{\mbox{d}s}{\mbox{d}s'}\equiv\lambda'(s')
\;\;. \end{equation}   
It can be shown that the corresponding infinitesimal tranformations actually generate the evolution 
of the system. This is the basis of statements that there is no time in systems   
where dynamics is pure gauge, i.e. of the ``problem of time''. 
We refer to \cite{ES02} for further discussion.    
  
Instead, with an evolution obviously taking place in such systems, 
we conclude from these remarks that the space-time description of motion requires 
a gauge invariant construction of a suitable time, replacing the fictitious proper time 
$\tau$. To this we add the important requirement that such construction should be based on 
quasi-local measurements, since global information (such as invariant path length)   
is generally not accessible to an observer in more realistic, typically nonlinear or higherdimensional theories.  

\subsection{``Timing'' Through an Extradimensional Window}   
Our construction of a physical time is based on the assumption that an observer 
in (3+1)-dimensional Minkowski 
space can perform measurements on full (5+1)-dimensional trajectories, 
however, only within a quasi-local window to the two extradimensions. 
In particular, the observer records the incidents 
(``units of change'') when the full trajectory hits an idealized detector which 
covers a small convex area element on the torus  
(compactified coordinates $x^{4,5}$).\footnote{One 
could invoke a popular distinction between brane and bulk matter 
as in string theory inspired higherdimensional cosmology, in order to construct more realistic 
models involving local interactions.} 

Thus, our aim is to construct time as an emergent   
quantity related to the increasing number of incidents measured by 
the reparametrization invariant incident number: 
\begin{equation}\label{incidentnumber}
I\equiv\int_{s_i}^{s_f}\mbox{d}s'\lambda (s')D(x^4(s'),x^5(s'))
\;\;, \end{equation} 
where $x^{4,5}$ describe the trajectory of the particle in the extradimensions, 
the integral is taken over the interval which corresponds to a 
given invariant path $x^\mu_i\rightarrow x^\mu_f$,  
and the function $D$ represents the detector features. 
Operationally it is not necessary to know the invariant path, in 
order to count the incidents.  

In the 
following examples we choose for $D$ the characteristic function 
of a small square of area $d^2$,  
$D(x^4,x^5)\equiv C_d(x^4)C_d(x^5)$, with $C_d(x)\equiv\Theta (x)(1-\Theta (x-d))$, 
which could be placed arbitrarily. Our 
results will not depend on the detailed shape of this idealized detector, if  
it is sufficiently small. More precisely, an incident is recorded only when, for 
example,  
the trajectory either leaves or enters the detector, or according to some other 
analogous restriction which could be incorporated into the 
definition of $D$. Furthermore, in order not to undo 
records, we have to restrict the lapse function $\lambda$ to be (strictly) 
positive, which also avoids trajectories which trace themselves backwards (or stall). 
The records correspond to  
a uniquely ordered series of events in Minkowski space, which are counted, and only 
their increasing total number is recorded, which is the Lorentz invariant 
incident number.    
  
Considering particularly the free motion on the torus, solution (\ref{xsolution}) yields:  
\begin{equation}\label{torusmotion}
[\vec\phi (\tau )]=[\vec\phi_0+\vec\pi\tau ] 
\;\;, \end{equation} 
where $\vec\phi$ is the vector formed of the angles $\phi^{4,5}$, and correspondingly 
$\vec\pi$, with 
$\pi^{4,5}\equiv p^{4,5}/2\pi R$; the quantities in \,(\ref{torusmotion}) are  
periodically continued, as before, see (\ref{compact1})-(\ref{compact2}). 
Without loss of generality we choose  
$\vec\phi_0=0$ and $\pi^{5}>\pi^{4}>0$, 
and place the detector next to the origin with 
edges aligned to the positive coordinate axes for simplicity. 

Since here we are not interested in what happens between the incidents, we reduce 
the description of the internal motion to coupled maps.  
For proper time intervals 
$\Delta\tau$ with $\pi^4\cdot\Delta\tau=1$, the $\phi^4$-motion is replaced by the map 
$m\longrightarrow m+1$, where $m$ is a nonnegative integer, while: 
\begin{equation}\label{phi5}  
[\phi^5]=[Pm]
\;\;, \end{equation} 
with $P\equiv\pi^5/\pi^4>1$. 
Then, also the detector response counting incidents can be 
represented as a map:  
\begin{equation}\label{I}
I(m+1)=I(m)+\Theta (\delta -[\phi^5])+\Theta ([\phi^5]-(1-P\delta ))
\;\;, \end{equation}
with $I(0)\equiv 1$, and where $\delta\equiv d/2\pi R$ corresponds to the detector edge length $d$, 
assumed to be sufficiently small, $P\delta\ll 1$. The two $\Theta$-function contributions 
account for the two different edges through which the trajectory can enter the 
detector in the present configuration. 

The nonlinear two-parameter map (\ref{I}) has surpring universal features, some of 
which we explored in \cite{E03}. Here, first of all, following the reparametrization invariant 
construction up to this point, we identify the {\it physical time} $T$ in terms  
of the incident number $I$ from (\ref{I}): 
\begin{equation}\label{time} 
T\equiv \frac{I}{\delta (\pi^4+\pi^5)}  
\;\;. \end{equation}   
A statistical argument for the scaling factor $\delta^{-1}(\pi^4+\pi^5)^{-1}$, based on  
ergodicity, has been 
given in \cite{ES02}, which applies here similarly. 

\begin{figure}[htb]
\begin{center}
\includegraphics[width=.8\textwidth]{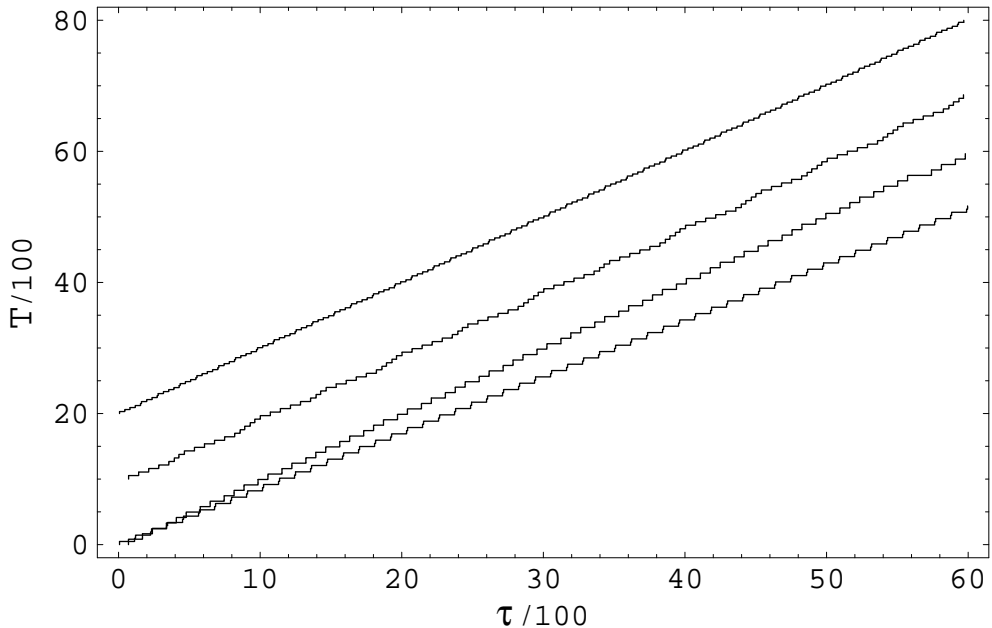}
\end{center}
\caption[]{The physical time $T$ as a function of proper time $\tau$ with detector parameter $\delta =.005$ and ratio of initial internal momenta $P=\sqrt{31}\;(\mbox{top}), e, \sqrt{2}, \pi\;(\mbox{bottom})$ (see main text); upper two curves displaced upwards by +10 and +20 units, respectively, for better visibility (from \cite{E03}).}
\end{figure}

We show in Fig.\,1  
how the physical time T typically is correlated with the fictitious proper time $\tau$. The  
proper time is extracted as those $m$-values when incidents happen: 
$\tau =m+1\Longleftrightarrow I(m+1)-I(m)=1$, corresponding to a particularly simple     
specification of the detector response. For a sufficiently small detector other such  
specifications yield the same results as described here. This is achieved by always rounding the  
extracted proper time values to integers, 
involving negligible errors of order $\delta,\delta /P\ll 1$. 

We find that the time $T$ does not run smoothly. This is due to the  
coarse-grained description of the internal motion: as if we were reading an analog clock 
under a stroboscopic light. In our construction, it is caused by the reduction of the full 
motion to a map (Poincar\'e section), corresponding to the recording of the physical incidents 
by the quasi-local detector. 

Furthermore, already after a short while, i.e. at low 
incident numbers, the constructed time approximates well the 
proper time $\tau$ on average.\footnote{The apparent excursion for the parameter value $P=\pi$ in Fig.\,1, 
does not persist for longer times, as shown in \cite{E03}.} 
The fluctuations on top of the observed linear dependence 
result in the {\it discreteness} of 
the constructed time. 

While in reference \cite{E03} we further analyzed the statistical properties of the map 
considered in this example, we take from here only the result that a physical time can be constructed 
based on suitable localized observations. Furthermore, after embarking on some useful formal developments in the 
next section, we will incorporate the resulting probabilistic mapping between discrete physical time and proper time 
of the equations of motion in Section\,4.  

\section{Classical Mechanics in Path-integral Form}   
Classical mechanics can be cast into path-integral form, as originally 
developed by Gozzi, Reuter and Thacker \cite{GRT}, and with recent addenda reported in \cite{GR00}.  
While the original motivation has been to provide a better understanding of geometrical aspects of quantization,   
we presently use it as a convenient tool. We refer the interested reader to the cited references for details,  
on the originally resulting extended (BRST type) symmetry in particular. We suitably incorporate 
time-reparametrization invariance, assuming equations of motion written 
in terms of proper time.
  
Let us begin with a $(2n)$-dimensional classical phase space ${\cal M}$ with coordinates denoted 
collectively by $\varphi^a\equiv (q^1,\dots ,q^n;p^1,\dots ,p^n),\;a=1,\dots ,2n$, where 
$q,p$ stand for the usual coordinates and conjugate momenta. Given the proper-time independent 
Hamiltonian $H(\varphi )$, the equations of motion are: 
\begin{equation}\label{eom}
\frac{\partial}{\partial\tau}\varphi^a=\omega^{ab}\frac{\partial}{\partial\varphi^b}H(\varphi ) 
\;\;, \end{equation} 
where $\omega^{ab}$ is the standard symplectic matrix and $\tau$ denotes the proper time; 
summation over indices appearing twice is understood.      
 
To the equation of motion we add the (weak) Hamiltonian constraint, $C_H\equiv H(\varphi )-\epsilon 
\simeq0$, with $\epsilon$ a suitably chosen parameter. This constraint has to be satisfied by the solutions 
of the equations of motion. Generally,  
it arises in reparametrization invariant models, similarly as the mass-shell constraint in the case  
of the relativistic particle \cite{E03}. It is necessary when the  
Lagrangian time parameter is replaced by the proper time in the equations 
of motion. In this way, an arbitrary ``lapse function'' is eliminated, which 
otherwise acts as a Lagrange multiplier for this constraint. 
 
We remark that field theories can be treated analogously, considering indices $a,b$, etc. 
as continuous variables.  
  
Starting point for our following considerations is the {\it classical} generating 
functional, 
\begin{equation}\label{Z} 
Z[J] \equiv \int_H{\cal D}\varphi\;\delta [\varphi^a(\tau )-\varphi^a_{cl} (\tau )]\exp (i\int\mbox{d}\tau\;J_a\varphi^a) 
\;\;, \end{equation} 
where $J\equiv\{ J_{a=1,\dots ,2n}\}$ is an arbitrary external source, 
$\delta [\cdot ]$ denotes a Dirac $\delta$-functional,  
and $\varphi_{cl}$ stands for a solution of the classical 
equations of motion satisfying the Hamiltonian constraint; its presence is indicated by the subscript ``$H$'' 
on the functional integral.  
The relevant boundary conditions shall be discussed in the following section. 
It is important to realize that $Z[0]$ gives weight 1 to a classical path satisfying the constraint 
and zero otherwise, integrating over all initial conditions.   
  
Using the functional equivalent of $\delta (f(x))=|\mbox{d}f/\mbox{d}x|_{x_0}^{-1}\cdot\delta (x-x_0)$\,,  
the $\delta$-functional under the integral for $Z$ can be replaced according to: 
\begin{equation}\label{delta} 
\delta [\varphi^a(\tau )-\varphi^a_{cl} (\tau )]=>
\delta [\partial_\tau\varphi^a-\omega^{ab}\partial_bH]
\det [\delta^a_b\partial_\tau -\omega^{ac}\partial_c\partial_bH] 
\;\;, \end{equation} 
slightly simplifying the notation, e.g. $\partial_b\equiv\partial /\partial\varphi^b$. 
Here the modulus of the functional determinant has been dropped \cite{GRT,GR00}.   

Finally, the $\delta$-functionals and determinant are  
exponentiated, using the functional Fourier representation and ghost variables, respectively. 
Thus, we obtain the generating functional in the convenient form: 
\begin{equation}\label{Zexp} 
Z[J]=\int_H{\cal D}\varphi{\cal D}\lambda{\cal D}c{\cal D}\bar c 
\;\exp \Big (i\int\mbox{d}\tau (L+J_a\varphi^a)\Big )
\;\;, \end{equation} 
which we abbreviate as 
$Z[J]=\int_H{\cal D}\Phi\;\exp (i\int\mbox{d}\tau L_J)$. 
The enlarged phase space is $(8n)$-dimensional, consisting of points described by the 
coordinates $(\varphi^a,\lambda_a,c^a,\bar c_a)$. The 
effective Lagrangian is now given by \cite{GRT,GR00}: 
\begin{equation}\label{L} 
L\equiv\lambda_a\Big (\partial_\tau\varphi^a-\omega^{ab}\partial_bH\Big ) 
+i\bar c_a\Big (\delta^a_b\partial_\tau -\omega^{ac}\partial_c\partial_bH\Big )c^b
\;\;, \end{equation} 
where $c^a,\bar c^a$ are anticommuting Grassmann variables. 
We remark that an entirely bosonic version of the path-integral exists \cite{GR00}.  

This completes our brief review of how to put (reparametrization invariant) 
classical mechanics into path-integral form.

\section{From Discrete Time to ``States''}   
We recall from our previous example that the discrete physical time $t$ 
has been obtained by counting suitably defined incidents, i.e., coincidences of  
points of the trajectory of the system with appropriate detectors \cite{ES02,E03}. 
Thus, it is given by a nonnegative integer multiple of some unit time, $t\equiv nT$. 
Then, we would like to express the proper time $\tau$ which parametrizes the 
evolution in terms of $t$. 

Here we assume instead that the physical time $t$ is mapped onto a normalized probability 
distribution $P$ of proper time values $\tau$: 
\begin{equation}\label{P} 
P(\tau ;t)\equiv\exp\Big (-S(\tau ;t)\Big )\;\;,\;\;\;\int\mbox{d}\tau\;P(\tau ;t)=1
\;\;. \end{equation} 
For uniqueness, we require that if 
$S(\tau ;t_1)$ and $S(\tau ;t_2)$, for $t_1\neq t_2$, have overlapping support, then 
they should coincide in this region.   

Thus, we describe the idealized case that the system can be separated into degrees 
of freedom which are employed in the construction of a physical ``clock'', yielding the 
values of $t$, and remaining degrees of freedom evolving in proper time. Neglecting  
the interaction between both components, and the details of the clock in particular, 
we describe the relation between physical and proper time by a probability distribution. 
In this situation, the Hamiltonian constraint only applies to the remaining degrees of 
freedom, while generally the system will be constrained as a whole.    
These aspects were exemplified in detail in the simple models of \cite{ES02,E03}. 

Correspondingly, we introduce the modified   
generating functional:   
\begin{equation}\label{Zdef}
Z[J]\equiv\int_H\mbox{d}\tau_i\mbox{d}\tau_f
\int{\cal D}\Phi\;\exp\Big (
i\int_{\tau_i}^{\tau_f}\mbox{d}\tau\; L_J-S(\tau_i;t_i)-S(\tau_f;t_f)\Big ) 
\;\;, \end{equation} 
instead of (\ref{Zexp}), using the condensed notation introduced there. In the present case, 
$Z[0]$ sums over all classical paths satisfying the 
constraint with weight $P(\tau_i;t_i)\cdot P(\tau_f;t_f)$, depending on their initial 
and final proper times, while all other paths get weight zero. 
In this way, 
the distributions of proper time values $\tau_{i,f}$ associated with the initial and 
final physical times, $t_i$ and $t_f$, respectively, are incorporated. 

Next, we insert $1=\int\mbox{d}\tau P(\tau ;t)$ into the expression for $Z$, 
with an arbitrarily chosen physical time $t>t_0$, and with $t_i=t_f\equiv t_0$. 
We require the two sets of trajectories created in this way to present branches of forward (``$>$'') 
and backward (``$<$'') motion.   
This leads us to factorize 
the path-integral into two connected ones: 
\begin{eqnarray} 
Z[J]=\int\mbox{d}\tau\;P(\tau ;t)
&\cdot&\int\mbox{d}\tau_f\int_H{\cal D}\Phi_<\;\exp\Big (
i\int_{\tau}^{\tau_f}\mbox{d}\tau'\; L_J^<-S(\tau_f;t_0)\Big ) 
\nonumber \\ [1ex] 
&\cdot&\int\mbox{d}\tau_i\int_H{\cal D}\Phi_>\;\exp\Big (
i\int_{\tau_i}^{\tau }\mbox{d}\tau''\; L_J^>-S(\tau_i;t_0)\Big ) 
\nonumber \\ [1ex]\label{Zfactors}
&\cdot&\prod_a\delta(\varphi^a_>(\tau )-\epsilon (a)\varphi^a_<(\tau ))
\;\;,  \end{eqnarray} 
where $\epsilon (a\leq n)\equiv 1,\;\epsilon (a\geq n+1)\equiv -1$\,, and 
$J\equiv J_>,J_<$\,, depending on the branch. 
The ordinary $\delta$-functions assure 
continuity of the classical paths in terms of the coordinates $q^a,\;a=1,\dots ,n$,  
and reflect the momenta $p^a,\;a=1,\dots ,n$, at proper time $\tau$. 

We observe that the generating functional will only be 
independent of the physical time $t$, in the absence of an external source, 
if we assume that the probability distribution $P$ does not explicitly depend on time, 
$-\log P(\tau ;t)=S(\tau ;t)\equiv S(\tau -t)$, and if we suitably specify the {\it boundary conditions}. 
We set: 
\begin{equation}\label{bc} 
\varphi^a_>(\tau_i)=\epsilon (a)\varphi^a_<(\tau_f)\;\equiv\;\phi^a(t_0)\;,\;\;a=1,\dots 2n
\;\;. \end{equation}  
Note that the boundary conditions are defined at the physical time $t_0$, to 
which correspond the distributed values of the proper times $\tau_{i,f}$. 
This establishes a one-to-one correspondence between both sets of trajectories. 
They could be viewed as closed loops with reflecting boundary conditions at both ends, 
$t_0$ and $t$, and fixed initial condition at $t_0$.  

Exponentiating the $\delta$-functions via Fourier transformation, the 
generating functional can be recognized indeed as a scalar product of a ``state'' and 
its adjoint.  
We define the normalized states by the path-integral: 
\begin{equation}\label{state} 
|\tau ,\pi_a;t\rangle\equiv Z[J]^{-1/2}
\int\mbox{d}\tau_i\int_H{\cal D}\Phi\;\exp\Big (
i\int_{\tau_i}^{\tau +t}\mbox{d}\tau'\; L_J
-S(\tau_i;t_0)+i\pi_a\varphi^a(\tau +t)\Big ) 
, \end{equation} 
and, similarly, the adjoint states: 
\begin{equation}\label{adjstate} 
\langle\tau ,\pi_a;t|\equiv Z[J]^{-1/2} 
\int\mbox{d}\tau_f\int_H{\cal D}\Phi\;\exp\Big (
i\int_{\tau +t}^{\tau_f}\mbox{d}\tau'\; L_J
-S(\tau_f;t_0)-i\pi_a\varphi^a(\tau +t)\Big ) 
, \end{equation} 
where the paths are forward and backward going as indicated by the integral 
boundaries in the exponent respectively, dropping ``$>,<$''; note that  
the summation is to be read as 
$\sum_a\epsilon(a)\pi_a\varphi^a$ in (\ref{adjstate}). 

The redundancy in designating the states, which depend on the sum 
of proper and physical time, only arises here, since the probability distribution $P$ is 
assumed not to be explicitly depending on the physical time, for simplicity. 

The scalar product of two such states is now defined, and calculated, as follows: 
\begin{equation}\label{overlap}
\langle t_2|t_1\rangle\equiv
\int\mbox{d}\tau\mbox{d}\pi\;P(\tau )\langle\tau ,\pi;t_2|\tau ,\pi;t_1\rangle =
\delta_{t_2,t_1}  
\;\;, \end{equation}
with $\mbox{d}\pi\equiv\prod_a(\mbox{d}\pi_a/2\pi )$. 
In particular, we have $\langle t|t\rangle =1$, which corresponds to (\ref{Zfactors}), 
using definitions (\ref{state}) and (\ref{adjstate}). Furthermore, for 
$t_1\neq t_2$, we find that states are orthogonal, $\langle t_2|t_1\rangle =0$, 
by the symmetry of the motion on the forward and backward branches, for a correspondingly 
symmetric source $J$, and by uniqueness 
of the Hamiltonian flow generating the paths.  

A remark is in order here concerning the integration over $\mbox{d}\pi$ above, which originates from 
exponentiating the $\delta$-functions of (\ref{Zfactors}). Due to the presence of the Hamiltonian 
constraints on both branches of a trajectory, one of the $\delta$-functions is redundant. We absorb the  
resulting $\delta (0)$ in the normalization of the states. 
  
Finally, the symmetry between the states and the adjoint states, given 
the stated assumptions, is perfect. We find: 
\begin{equation}\label{ccstates} 
\langle\tau ,\pi ;t|=|\tau ,\pi ;t\rangle^\ast   
\;\;, \end{equation} 
which is a desirable property of states in a Hilbert space 
(in ``$\tau ,\pi$-representation''). 
However, from our heuristic discussion this appears as a restriction which could be relaxed, 
resulting in a less familiar 
relation between the vector space and its dual.\footnote{One might consider only 
forward going paths, for example, with boundary conditions on the coordinates set at $\pm t_0$. 
In this case, however, it is not obvious how to obtain the correspondent of (\ref{ccstates}).} 

\section{Unitary Evolution}   
Following the same approach which led to the definition of states in 
(\ref{state}),(\ref{adjstate}), we consider the time evolution of states 
in the absence of a source, $J=0$. 
Suitably inserting ``1'', as before, and splitting the path-integral, we obtain:  
\begin{equation}\label{evol} 
|\tau',\pi';t'\rangle=\int\mbox{d}\tau\mbox{d}\pi\;P(\tau) 
U(\tau',\pi';t'|\tau ,\pi ;t)|\tau ,\pi ;t\rangle
\;\;, \end{equation}
with the kernel:    
\begin{equation}\label{kernel} 
U(\tau',\pi';t'|\tau ,\pi ;t)
\equiv\int{\cal D}\Phi\;\exp\Big (
i\int_{\tau +t}^{\tau'+t'}\mbox{d}\tau''\; L
+i\pi'\cdot\varphi (\tau'+t')
-i\pi\cdot\varphi (\tau +t) 
\Big ) 
, \end{equation}
where the integral is over {\it all paths} running between $\tau +t$ and $\tau'+t'$, 
subject to the constraint; 
here we abbreviate $\pi\cdot\varphi\equiv\pi_a\varphi^a$.  
We interpret this as a matrix element of the evolution operator $\widehat U(t'|t)$.  
  
Then, it is straightforward to establish the following composition rule: 
\begin{equation}\label{composition}
\widehat U(t''|t')\cdot\widehat U(t'|t)=\widehat U(t''|t) 
\;\;, \end{equation} 
where integration over the intermediate variables, say $\tau',\pi'$\,, 
with appropriate weight factor $P(\tau')$, is understood. 
These integrations effectively remove the ``1'', which is inserted when 
factorizing path-integrals, and link the endpoint coordinates of one classical path to the 
initial of another. 

Since the Hamiltonian constraint is a constant of motion, there is no need to 
constrain the path integral representing the evolution operator. Integrating over 
intermediate variables removes all contributions violating the Hamiltonian 
constraint, provided we work with properly constrained states. This will be 
further discussed in the following section. 
    
The physical-time dependence of the evolution operator  
amounts to translations of proper time variables. Therefore, we may study further properties 
of $\widehat U$ without explicitly keeping it. This simplicity, of course, is related  
to the analogous property of the states, which we mentioned. 

We begin by rewriting the functional integral of (\ref{kernel}):   
\begin{equation}\label{Ut1}
U(\tau',\pi';\tau ,\pi )
=\int{\cal D}\varphi\;\delta [\varphi^a(\tau )-\varphi^a_{cl}(\tau )]
\exp\Big (i\pi'\cdot\varphi (\tau')-i\pi\cdot\varphi (\tau)\Big )
\;\;, \end{equation} 
cf. Section\,3, where the paths run between $\tau$ and $\tau'$, integrating over all 
initial conditions. Fixing the initial condition of a classical path, we can pull the exponential 
factors out of the integral, due to the $\delta$-functional,  
and integrate over all initial conditions in the end: 
\begin{equation}\label{Ut2} 
U(\tau',\pi';\tau ,\pi )
=\int\mbox{d}\varphi_i(\tau )\;
\exp\Big (i\pi'\cdot\varphi_f(\tau')-i\pi\cdot\varphi_i(\tau )\Big )
\int{\cal D}\varphi\;\delta [\varphi^a(\tau )-\varphi^a_{cl}(\tau )]
, \end{equation} 
where $\varphi_f(\tau')$ denotes the endpoint of the path singled out by the particular   
initial condition, $\varphi_i(\tau )$. The functional integral 
equals one. Then, we obtain the simple but central result: 
\begin{eqnarray}\label{Ut3}
U(\tau',\pi';\tau ,\pi )&=&
\int\mbox{d}\varphi\; 
\exp \Big (i\pi'\exp [\widehat{\cal L}(\tau'-\tau )]\cdot\varphi -i\pi\cdot\varphi\Big )
\\ [1ex] \label{Ut4} 
&\equiv&{\cal E}(\pi',\pi ;\tau'-\tau ) 
\;\;, \end{eqnarray} 
where $\mbox{d}\varphi\equiv\prod_a\mbox{d}\varphi^a$, and  
with the Liouville operator: 
\begin{equation}\label{Liouville}
\widehat{\cal L}\equiv -\frac{\partial H}{\partial\varphi}\cdot\omega\cdot
\frac{\partial}{\partial\varphi}
\;\;, \end{equation} 
which is employed in order to propagate the 
classical solution from the initial condition at $\tau$ to proper time $\tau'$; 
$\omega$ is the symplectic matrix. 
  
Using (\ref{Ut3}), one readily confirms (\ref{composition}) once again. 
In particular, then $\widehat U(t|t')\cdot\widehat U(t'|t)=\widehat U(t|t)$, 
which is not diagonal, in general, in this $\tau,\pi$-representation. We have:  
$U(\tau',\pi';t|\tau ,\pi ;t)={\cal E}(\pi',\pi ;\tau'-\tau )$, as defined 
in (\ref{Ut4}).  
  
In order to proceed, we consider the time dependence of the evolution kernel ${\cal E}$. 
It is determined by the equation: 
\begin{eqnarray}
i\partial_\tau {\cal E}(\pi',\pi ;\tau )&=&-\int\mbox{d}\varphi\; 
\exp \Big (i\pi'\cdot\varphi (\tau )-\pi\cdot\varphi\Big )\;\pi'\cdot\omega\cdot
\frac{\partial}{\partial\varphi}H(\varphi (\tau )) 
\nonumber \\ [1ex] \label{Et} 
&=&\widehat{\cal H}(\pi',-i\partial_{\pi'}){\cal E}(\pi',\pi ;\tau )
\;\;, \end{eqnarray} 
with the effective {\it Hamilton operator}: 
\begin{equation}\label{Hamiltonian} 
\widehat{\cal H}(\pi ,-i\partial_\pi )\equiv -\pi \cdot\omega\cdot
\frac{\partial}{\partial\varphi}H(\varphi )|_{\varphi =-i\partial_{\pi}}  
\;\;. \end{equation} 
Here we used (\ref{Ut3})--(\ref{Ut4}), together with the equation of motion 
(\ref{eom}). The initial condition is:  
\begin{equation}\label{Einitial} 
{\cal E}(\pi',\pi ;0)=(2\pi )^{2n}\delta^{2n}(\pi'-\pi) 
\;\;, \end{equation} 
as read off from (\ref{Ut3}).   
  
Using (\ref{Et}), we finally obtain the {\it Schr\"odinger equation} which describes 
the evolution of the states in physical time: 
\begin{eqnarray}   
i\partial_t\langle\tau ,\pi |\Psi (t)\rangle &=& 
\int\mbox{d}\tau'\mbox{d}\pi'\;P(\tau')i\partial_t{\cal E}(\pi ,\pi';\tau +t-\tau')
\langle\tau',\pi'|\Psi (0)\rangle 
\nonumber \\ [1ex] \label{Schroedinger} 
&=&\widehat{\cal H}(\pi ,-i\partial_\pi )\langle\tau ,\pi |\Psi (t)\rangle 
\;\;. \end{eqnarray} 
Here we implicitly employed the relation $\langle\tau ,\pi|\Psi (t)\rangle 
=\langle\tau +t,\pi|\Psi (0)\rangle =\langle 0,\pi|\Psi (t+\tau )\rangle$, 
in order to analytically continue to real values of $t$ and to perform the derivative, 
despite that the physical time is discrete. 

Clearly, the hermitian Hamiltonian 
must be incorporated in a unitary transfer matrix, in order to describe the 
evolution through one discrete physical time step. It is plausible that presently the 
need for regularization of the Hamilton operator, demonstrated in subsequent sections, 
arises here. It is also conceivable that in a more realistic situation, with clock 
degrees of freedom forming dynamically part of the system, this particular complication 
is alleviated.  

Furthermore, considering stationary states, we have:
\begin{eqnarray}\label{stationary} 
\langle\tau ,\pi|\Psi_E(t)\rangle&\equiv&\exp (-iEt)\langle\tau ,\pi|\Psi (0)\rangle 
=\exp (-iE(t+\tau ))\langle 0,\pi|\Psi (0)\rangle
\\ [1ex] \label{stationary1}
&\equiv&\exp (-iE(t+\tau ))\langle \pi|\Psi_E\rangle
\;\;, \end{eqnarray} 
due to the previously discussed additivity of proper and physical 
time in the present context. Similarly, the Hamiltonian $\widehat{\cal H}$ is independent 
of the probability distribution $P$, mapping physical to proper time, since in the presently 
idealized situation the clock is decoupled from the system. 

Note that there is {\it no} $\hbar$ in our 
equations. If introduced, it would merely act as a conversion factor of units. 
On the other hand, there is an intrinsic scale corresponding to the clock's unit time interval $T$, 
which could be analyzed in a more complete treatment where clock and mechanical system are 
part of the Universe and interact. 
 
Before we will illustrate in some examples the type of quantum Hamiltonians that one obtains, 
we have to first address the classical observables and their place in the emergent 
quantum theory, in particular we need to implement the classical Hamiltonian constraint. 
We recall that in a reparametrization invariant 
classical theory the Hamiltonian constraint is an essential ingredient related to 
the gauge symmetry one is dealing with.     

\section{Observables}  
It follows from our introduction of states in Section\,4, see particularly (\ref{Zdef})--(\ref{adjstate}), 
how the classical observables of the underlying mechanical system can be determined. Considering observables 
which are function(al)s of the phase space variables $\varphi$, the definition of their expectation value 
at physical time $t$ is obvious: 
\begin{eqnarray}\label{expect1} 
\langle O[\varphi ];t\rangle&\equiv&\int\mbox{d}\tau\;P(\tau ;t)O[-i\frac{\delta}{\delta J(\tau )}]\log Z[J]|_{J=0}
\\ [1ex] \label{expect2}
&=&\int\mbox{d}\tau\mbox{d}\pi\;P(\tau -t)\langle\tau ,\pi ;0|O[\varphi (\tau )]|\tau ,\pi ;0\rangle 
\\ [1ex] \label{expect3} 
&=&\int\mbox{d}\tau\mbox{d}\pi\;P(\tau )\langle\tau ,\pi ;t|O[\varphi (\tau +t)]|\tau ,\pi ;t\rangle 
\\ [1ex] \label{expect4} 
&=&\int\mbox{d}\tau\mbox{d}\pi\;P(\tau )\langle\tau ,\pi ;t|O[-i\partial_\pi ]|\tau ,\pi ;t\rangle 
\\ [1ex] \label{expect5}
&=&\langle\Psi (t)|\widehat O[\varphi ]|\Psi (t)\rangle
\;\;, \end{eqnarray} 
where: 
\begin{equation}\label{phihat}
\widehat O[\varphi ]\equiv O[\widehat\varphi ]\;\;,\;\;\; 
\widehat\varphi\equiv -i\partial_\pi 
\;\;, \end{equation} 
in $\tau ,\pi$-representation. In (\ref{expect2})--(\ref{expect3}) 
the notation is symbolical, since the observable should be properly included 
in the functional integral defining the ket state, for example.  

Thus, a classical observable is represented by the corresponding function(al) of a suitably defined 
{\it momentum} operator. Furthermore, its expectation value at physical time $t$ is represented by the  
effective quantum mechanical expectation value of the corresponding operator with respect to the physical-time 
dependent state under consideration, which incorporates the 
weighted average over the proper times $\tau$, according to the distribution $P$. Not quite surprisingly, 
the evaluation of expectation values involves an integration over the whole $\tau$-parametrized 
``history'' of the states.

Furthermore, making use of the evolution operator $\widehat U$ of Section\,5, in order to refer observables at different 
proper times $\tau_1,\tau_2,\dots$ to a common reference point $\tau$, one can construct {\it correlation functions} of 
observables as well, similarly as in \cite{Wetterich02}, for example.   

The most important observable for our present purposes is the classical Hamiltonian, $H(\varphi )$, 
which enters the Hamiltonian constraint of a classical reparametrization invariant system. It is, by 
assumption, a constant of the classical motion. However, it is easy to see that also its quantum 
descendant, $\widehat H(\varphi )\equiv H(\widehat\varphi )$, is conserved, since it commutes 
with the effective Hamiltonian of (\ref{Hamiltonian}): 
\begin{eqnarray}\label{Hconserv} 
[\widehat H,\widehat{\cal H}]&=& 
H(-i\partial_\pi )\;\pi \cdot\omega\cdot
\frac{\partial}{\partial\varphi}H(\varphi )|_{\varphi =-i\partial_{\pi}}
-\pi \cdot\omega\cdot
\frac{\partial}{\partial\varphi}H(\varphi )|_{\varphi =-i\partial_{\pi}}\;H(-i\partial_\pi )
\nonumber \\ [1ex] \label{Hconserv1} 
&=&\frac{\partial}{\partial\varphi}H(\varphi )|_{\varphi =-i\partial_{\pi}} 
\cdot\omega\cdot
\frac{\partial}{\partial\varphi}H(\varphi )|_{\varphi =-i\partial_{\pi}}\;=\;0 
\;\;, \end{eqnarray} 
due to the antisymmetric character of the symplectic matrix. 
Therefore, it suffices to implement the Hamiltonian constraint at an arbitrary  
time. 
  
Then, the constraint of the form $C_H\equiv H[\varphi ]-\epsilon\simeq 0$ may be  
incorporated into the definition of the states in (\ref{state}) 
by including an extra factor $\delta (C_H)$ into the functional integral, and 
analogously for the adjoint states. Exponentiating the $\delta$-function, we  
pull the exponential out of the functional integral, as before. Thus, we  
find the following operator representing the constraint: 
\begin{equation}\label{constraint} 
\widehat{C}\equiv\int\mbox{d}\lambda\;\exp\Big (i\lambda (\widehat H(\varphi )-\epsilon )\Big )
=\delta (\widehat{C}_H)
\;\;, \end{equation}
which acts on states as a projector. 
Of course, a corresponding number of projectors 
should be included into the definition of the generating functional, see (\ref{Zfactors}), 
for appropriate normalization of the states. 
  
Supplementing (\ref{expect1})--(\ref{expect5}) by the 
insertion of the Hamiltonian constraint, the properly constrained expection 
values of observables should be calculated according to: 
\begin{equation}\label{Cexpect}  
\langle O[\varphi ];t\rangle_H\equiv\langle\Psi (t)|\widehat O[\varphi ]\widehat{C}|\Psi (t)\rangle
\;\;, \end{equation}
which will deviate from the results of the previous definition.

Finally, also the eigenvalue problem of stationary 
states, see (\ref{stationary})--(\ref{stationary1}), 
should be studied in the projected subspace: 
\begin{equation}\label{eigenvalue}  
\widehat{\cal H}\widehat{C}|\Psi\rangle =E\widehat{C}|\Psi\rangle 
\;\;, \end{equation} 
to which we shall return in the following examples. 

\section{Examples of Emergent Quantum Systems}   
\subsection{Quantum Harmonic Oscillator from Classical One Beneath}  
All {\it integrable models} can be presented as collections of  
independent harmonic oscillators. Therefore, we begin with  
the harmonic oscillator of unit mass and of frequency $\Omega$. The action is: 
\begin{equation}\label{oscaction} 
S\equiv\int\mbox{d}t\;\Big (\frac{1}{2\lambda}(\partial_tq)^2-\frac{\lambda\Omega^2}{2}(q^2-2\epsilon)\Big ) 
\;\;, \end{equation} 
where $\lambda$ denotes the arbitrary lapse function, i.e. Lagrange multiplier for the Hamiltonian constraint, and $\epsilon >0$ is the parameter fixing the energy presented by this constraint. 
   
Introducing the proper time, $\tau\equiv\int\mbox{d}t\;\lambda$, the Hamiltonian equations of 
motion and Hamiltonian constraint for the oscillator are:  
\begin{eqnarray}\label{osceom1} 
\partial_\tau q=p\;\;,\;\;\;\partial_\tau p=-\Omega^2q\;\;, 
\\ [1ex] \label{oscconstraint} 
\frac{1}{2}(p^2+\Omega^2q^2)-\epsilon =0
\;\;, \end{eqnarray} 
respectively.  
  
Comparing the general structure of the equations of motion (\ref{eom}) with 
the ones obtained here, we identify the effective Hamilton operator (\ref{Hamiltonian}), 
while the constraint operator follows from (\ref{constraint}): 
\begin{eqnarray}\label{oscHamiltonian}
&\;&\widehat{\cal H}=-(\pi_q\widehat\varphi_p-\Omega^2\pi_p\widehat\varphi_q)= 
-\pi_q(-i\partial_{\pi_p})+\Omega^2\pi_p(-i\partial_{\pi_q})\;\;, 
\\ [1ex] \label{osccoperator} 
&\;&\widehat{C}
=\delta (\widehat\varphi_p^{\;2}+\Omega^2\widehat\varphi_q^{\;2}-2\epsilon )=
\delta (\partial_{\pi_p}^{\;2}+\Omega^2\partial_{\pi_q}^{\;2}+2\epsilon)
\;\;, \end{eqnarray}
respectively. Here we employ the convenient notation 
$\varphi^a\equiv (\varphi_q;\varphi_p)$, and 
correspondingly $\pi^a\equiv (\pi_q;\pi_p),\;\partial_\pi^a\equiv (\partial_{\pi_q};\partial_{\pi_p})$.    
Further simplifying this with the help of polar coordinates, 
$\pi_q\equiv -\Omega\rho\cos\phi$ and $\pi_p\equiv\rho\sin\phi$, we obtain: 
\begin{eqnarray}\label{angoscHamiltonian}
&\;&\widehat{\cal H}
=\Omega\widehat L_z=-i\Omega\partial_\phi\;\;, 
\\ [1ex] \label{osccoperator1} 
&\;&\widehat{C}
=\delta (\Delta_2+2\epsilon)
=\delta (\partial_\rho^{\;2}+\rho^{-1}\partial_\rho +\rho^{-2}\partial_\phi^{\;2}+2\epsilon )
\;\;, \end{eqnarray}
where $\widehat L_z$ denotes the $z$-component of the usual angular momentum 
operator and $\Delta_2$ the Laplacian in two dimensions. 

We observe that the eigenfunctions of the eigenvalue problem 
posed here factorize into a radial and an angular part. The radial eigenfunction, a  
cylinder function,  is important for the calculation of expectation values of 
certain operators and the overall normalization of the resulting wave functions. 
However, it does not influence the most interesting spectrum of the Hamiltonian.     
  
In the absence of the full angular momentum algebra, we discretize the angular derivative. 
Then, the energy eigenvalue problem consists in: 
\begin{equation}\label{osceigen}
\widehat{\cal H}\psi (\phi_n) 
=-i(\Omega N/2\pi )\Big (\psi (\phi_{n+1})-\psi (\phi_n)\Big ) =E\psi (\phi_n) 
\;\;, \end{equation} 
with $\phi_n\equiv 2\pi n/N$, $1\leq n\leq N$, and the continuum limit will be 
considered momentarily.   
  
A complete orthonormal set of eigenfunctions and the eigenvalues are: 
\begin{eqnarray}\label{osceigenfunction} 
\psi_m(\phi_n)&=&N^{-1/2}\exp [i(m+\delta )\phi_n]\;\;,\;\;1\leq m\leq N  
\;\;, \\ [1ex] \label{osceigenvalue} 
E_m&=&i(\Omega N/2\pi )\Big (1-\exp [2\pi i(m+\delta)/N]\Big )
\\ [1ex] \label{osceigenvalue1}
&\stackrel{N\rightarrow\infty}{\longrightarrow}&\Omega (m+\delta )\;\;,\;\;m\in\mathbf{N}  
\;\;, \end{eqnarray} 
where $\delta$ is an arbitrary real constant.   

Obviously, the freedom in choosing the constant $\delta$, which arises from 
the regularization of the Hamilton operator, is very wellcome. Choosing $\delta\equiv -1/2$,  
we arrive at the {\it quantum harmonic oscillator}, starting from the corresponding 
classical system. Thus, we recover in a straightforward way 't Hooft's result, derived from an 
equivalent cellular automaton \cite{tHooft01}. See also \cite{E03} for the  
completion of a similar quantum model. In the following example we will encounter 
one more model of this kind and demonstrate its solution in detail.   

Here, and similarly in following examples, the {\it eigenvalues are complex}, with the real 
spectrum only obtained in the continuum limit. This is due to the fact that we discretize 
first-order derivatives most simply, i.e. asymmetrically. It can be avoided easily by 
employing a symmetric discretization, if necessary. 

We find it interesting that our general Hamilton operator (\ref{Hamiltonian})  
does not allow for the direct addition of a constant energy term, while the regularization 
performed here does.     

\subsection{Quantum System with Classical Relativistic Particle Beneath}  
Introducing proper time as in Section\,2, but leaving the extradimensions for now, 
the equations of motion and 
the Hamiltonian constraint of the reparametrization invariant kinematics of a 
classical relativistic particle of mass $m$ are given by: 
\begin{eqnarray}\label{releom}
\partial_\tau q^\mu =m^{-1}p^\mu\;\;,\;\;\;\partial_\tau p^\mu =0
\;\;, \\ [1ex] \label{relc} 
p\cdot p-m^2=0 
\;\;, \end{eqnarray} 
respectively. Here we have $\varphi^a\equiv (q^0,\dots ,q^3;p^0,\dots ,p^3),\;a=1,\dots ,8$; 
four-vector products involve the Minkowski metric, $g_{\mu\nu}\equiv\mbox{diag}(1,-1,-1,-1)$. 
   
Proceeding as before, we identify the effective Hamilton operator: 
\begin{equation}\label{relHamiltonian}
\widehat{\cal H}=-m^{-1}\pi_q\cdot\widehat\varphi_p=-m^{-1}\pi_q\cdot(-i\partial_{\pi_p}) 
\;\;, \end{equation} 
corresponding to (\ref{Hamiltonian}); the notation is as   
introduced after (\ref{osccoperator}), however, involving four-vectors.  
Furthermore, the Hamiltonian constraint 
is represented by the operator: 
\begin{equation}\label{relconstraint} 
\widehat{C}
=\delta (\widehat\varphi_p^{\;2}-m^2)=\delta (\partial_{\pi_p}^{\;2}+m^2)
\;\;, \end{equation}
following from (\ref{constraint}). 

After a Fourier transformation, which replaces the variable $\pi_q$ by a derivative (four-vector) $+i\partial_x$, 
and with $\pi_p\equiv\bar x$, the Hamiltonian and constraint operators become: 
\begin{equation}\label{Hplusc}  
\widehat{\cal H}=-m^{-1}\partial_x\cdot\partial_{\bar x}\;\;,\;\;\;
\widehat{C}=\delta (\partial_{\bar x}^{\;2}+m^2) 
\;\;, \end{equation} 
respectively. 

Similarly as in the harmonic oscillator case, the eigenvalue problem is properly defined by 
solved by discretizing the system on a hypercubic lattice of volume $L^8$ 
(lattice spacing $l\equiv L/N$) 
with periodic boundary conditions, 
for example. Here we obtain the eigenfunctions:
\begin{equation}\label{releigenfunction}   
\psi_{k_x,k_{\bar x}}(x_n,\bar x_n)=N^{-1}
\exp [i(k_x+\delta_x)\cdot x_n+i(k_{\bar x}+\delta_{\bar x})\cdot \bar x_n] 
\;\;, \end{equation} 
with coordinates $x_n^\mu\equiv ln^\mu$ and   
momenta $k_x^\mu\equiv 2\pi k^\mu /L$, with $1\leq n^\mu ,k^\mu\leq N$, 
and where $\delta_x^\mu$ are arbitrary real constants,  
for all $\mu =0,\dots ,3$ (analogously $\bar x_n^\mu$, $k_{\bar x}^\mu$, 
$\delta_{\bar x}^\mu$). 

The energy eigenvalues are: 
\begin{eqnarray}\label{releigenvalue}  
E_{k_x,k_{\bar x}}&=&
-m^{-1}l^{-2}\Big ((\exp [il(k_x+\delta_x)^0]-1)(\exp [il(k_{\bar x}+\delta_{\bar x})^0]-1)
\\ 
&\;&\;\;\;\;\;\;\;\;\;\;-\sum_{j=1}^3(\exp [il(k_x+\delta_x)^j]-1)(\exp [il(k_{\bar x}+\delta_{\bar x})^j]-1)\Big )
\nonumber \\ [1ex] \label{releigenvalue1}
&=&
m^{-1}(k_x+\delta_x)\cdot (k_{\bar x}+\delta_{\bar x})+O(l)
\;\;, \end{eqnarray} 
where is $L$ is kept constant in the continuum limit, $l\rightarrow 0$. Furthermore, in this limit, 
one finds that the Hamiltonian constraint requires timelike ``on-shell'' vectors 
$k_{\bar x}$, obeying $(k_{\bar x}+\delta_{\bar x})^2=m^2$, 
while leaving $k_x$ unconstrained.   
  
Continuing, we perform also the infinite volume limit, $L\rightarrow\infty$, 
which results in a continuous energy spectrum in (\ref{releigenvalue1}). 
We observe that no matter how we choose the constants $\delta_x,\delta_{\bar x}$, 
the spectrum will not be positive definite. Thus, the emergent model is not 
acceptable, since it does not lead to a stable groundstate. 
  
However, let us proceed more carefully with the various limits involved 
and show that indeed a well-defined quantum model can be obtained.  
For simplicity, considering (1+1)-dimensional Minkowski space and anticipating 
the massless limit, we rewrite (\ref{releigenvalue1}) explicitly: 
\begin{equation}\label{releigen}
E_{k,\bar k}=-(\frac{2\pi}{\sqrt mL})^2
(\bar k^1+\bar\delta^1)\Big ((k^0+\delta^0)+(k^1+\delta^1)\Big )+O(m)
\;\;, \end{equation} 
where we suitably rescaled and renamed the constants and the momenta, which run in 
the range $1\leq\bar k^1,k^{0,1}\leq N\equiv 2s+1$. Furthermore, we incorporated the 
Hamiltonian (on-shell) constraint, such 
that only the positive root contributes: 
$\bar k^0+\bar\delta^0=|\bar k^1+\bar\delta^1|+O(m^2)=-(\bar k^1+\bar\delta^1)+O(m^2)$. 
This can be achieved by suitably choosing $\bar\delta^{0,1}$.     

In fact, just as in the previous harmonic oscillator case, 
the choice of the constants is crucial in defining the quantum model. 
Here we set: 
\begin{equation}\label{deltas}
\bar\delta^0\equiv\frac{1}{2}\;,\;\;\bar\delta^1\equiv \frac{1}{2}-2s-3\;,\;\;
\delta^{0,1}\equiv 0
\;\;. \end{equation} 
This results in the manifestly positive definite spectrum: 
\begin{equation}\label{releigen1}
E(\bar s_z,s_z^{0,1})=(\frac{2\pi}{\sqrt mL})^2\Big ((\bar s_z+s+\frac{1}{2})+1\Big )  
\Big ((s_z^0+s+\frac{1}{2})+(s_z^1+s+\frac{1}{2})+1\Big )+O(m)
\;\;, \end{equation}
with (half)integer quantum numbers $\bar s_z,s_z^{0,1}$,  
all in the range $-s\leq s_z\leq s$, replacing $\bar k^1,k^{0,1}$. 
  
Recalling the algebra of the $SU(2)$ generators, with $S_z|s_z\rangle =s_z|s_z\rangle$ 
in particular, we are led to consider the generic operator: 
\begin{equation}\label{Hdiag} 
h\equiv S_z+s+\frac{1}{2}   
\;\;, \end{equation}
i.e., diagonal with respect to $|s_z\rangle$-states of the (half)integer representations determined by $s$.  
In terms of such operators, we obtain the regularized Hamiltonian corresponding to (\ref{releigen1}): 
\begin{equation}\label{relHreg}
\widehat{\cal H}=
(\frac{2\pi}{\sqrt mL})^2\Big (1+\bar h+h_0+h_1+\bar h(h_0+h_1)\Big )+O(m)
\;\;, \end{equation}
which will turn out to be equivalent to three harmonic oscillators, including a coupling term 
plus an additional contribution to the vacuum energy. 

A Hamiltonian of the type of h has been the starting point of 't Hooft's analysis \cite{tHooft01}, 
which we adapt for our purposes in the following.  

Continuing with standard notation, we have $S^2\equiv S_x^{\;2}+S_y^{\;2}+S_z^{\;2}=s(s+1)$, which suffices 
to obtain the following identity: 
\begin{equation}\label{Hsq} 
h=\frac{1}{2s+1}\Big (S_x^{\;2}+S_y^{\;2}+\frac{1}{4}+h^2\Big ) 
\;\;. \end{equation} 
Furthermore, using $S_\pm\equiv S_x\pm iS_y$, we define coordinate and conjugate 
momentum operators: 
\begin{equation}\label{qp}
\hat q\equiv\frac{1}{2}(aS_-+a^\ast S_+)\;\;,\;\;\; 
\hat p\equiv\frac{1}{2}(bS_-+b^\ast S_+)
\;\;, \end{equation} 
where $a$ and $b$ are complex coefficients. Calculating the basic commutator with the help 
of $[S_+,S_-]=2S_z$ and using (\ref{Hdiag}), we obtain: 
\begin{equation}\label{commutator} 
[\hat q,\hat p]=i(1-\frac{2}{2s+1}h)
\;\;, \end{equation} 
provided we set $\Im (a^\ast b)\equiv -2/(2s+1)$. Incorporating this,  
we calculate: 
\begin{equation}\label{sumsqu}
S_x^{\;2}+S_y^{\;2}=\frac{(2s+1)^2}{4}\left (
|a|^2\hat p^2+|b|^2\hat q^2-(\Im a\cdot\Im b+\Re a\cdot\Re b)\{ \hat q,\hat p\}\right )
\;\;. \end{equation}  
In order to obtain a reasonable Hamiltonian in the continuum limit,  
we set: 
\begin{equation}\label{ab} 
a\equiv i\frac{\Omega^{-1/2}}{\sqrt{s+1/2}}\;\;,\;\;\;b\equiv\frac{\Omega^{1/2}}{\sqrt{s+1/2}} 
\;\;,\;\;\;\Omega\equiv (\frac{2\pi}{\sqrt mL})^2
\;\;. \end{equation}  
Then, the previous (\ref{Hsq}) becomes: 
\begin{equation}\label{Hsq1} 
\Omega h=\frac{1}{2}\hat p^2+\frac{1}{2}\Omega^2\hat q^2
+\frac{1}{(2s+1)\Omega}\Big (\frac{1}{4}\Omega^2+(\Omega h)^2\Big )
\;\;, \end{equation}
reveiling a nonlinearly modified harmonic oscillator Hamiltonian, similarly as in \cite{E03,tHooft01}. 

Now it is safe to consider the continuum limit, $2s+1=N\rightarrow\infty$, 
keeping $\sqrt mL$ and $\Omega$ finite. This produces the usual 
$\hat q,\hat p$-commutator in (\ref{commutator}) for states with limited energy 
and the standard harmonic oscillator Hamiltonian in (\ref{Hsq1}). 

Using these results in (\ref{relHreg}), the Hamilton operator of the  
emergent quantum model is obtained:  
\begin{equation}\label{relHreg1}
\widehat{\cal H}=\Omega +\frac{1}{2}\sum_{j=\bar 1,0,1}\Big (\hat p_j^{\;2}+\Omega^2\hat q_j^{\;2}\Big )
+\frac{1}{4\Omega}(\hat p_{\bar 1}^{\;2}+\Omega^2\hat q_{\bar 1}^{\;2})
\sum_{j=0,1}\Big (\hat p_j^{\;2}+\Omega^2\hat q_j^{\;2}\Big )
\;, \end{equation}
where the massless limit together with the infinite volume limit is carried out, 
$m\rightarrow 0,\;L\rightarrow\infty$, in such a way that $\Omega$ remains finite.  

The resulting Hamiltonian here is well defined in terms of continuous operators 
$\hat q$ and $\hat p$, as usual, and has a positive definite spectrum. The coupling term might 
appear slightly less unfamiliar, if the oscillator algebra is realized in terms of 
bosonic creation and annihilation operators. 

We previously calculated 
the matrix elements of operators $\hat q,\hat p$ with respect to the SU(2) basis 
of primordial states in a similar case, showing that localization of the quantum 
oscillator has little to do with localization in the classical model beneath \cite{ES02,E03} . 

Finally, we remark that had we chosen $\bar\delta^{0,1}=\delta^{0,1}\equiv 1/2$, 
instead of (\ref{deltas}), 
then a relative sign between terms would remain, originating from the 
Minkowski metric, and this would yield the Hamiltonian 
$\widehat{\cal H}\propto (1+\bar h)(h_0-h_1)$, which is not 
positive definite. Similarly, any 
symmetric choice, $\bar\delta^{0,1}=\delta^{0,1}\equiv\delta$ would 
suffer from this problem.     
 
This raises the important 
issue of the role of canonical transformations, and of symmetries in particular.  
It is conceivable that symmetries will play a role in restricting the present  
arbitrariness of the regularization defining a quantum model. 
We will address further aspects of this in the following section.  

\section{Remarks on (Non)Integrable Interactions}   
We resume our discussion of general features of the emergent quantum 
mechanics. Specifically, let us consider a classical system with $n$ 
degrees of freedom, for example, a chain of particles 
with harmonic coupling and anharmonic potentials. Denoting the phase space variables by 
$\varphi^a\equiv (Q,P)$, where $Q,P$ are $n$-component vectors, we assume for definiteness   
a Hamiltonian of the form, $H(\varphi )\equiv (1/2)P^2+V(Q)$, 
i.e. with a kinetic term which is simply quadratic in the momenta.   
  
In this case, following (\ref{Hamiltonian}) and (\ref{constraint}), and with: 
\begin{equation}\label{operators} 
\widehat Q\equiv -i\partial_{\pi_Q}\;\;, \;\;\;\widehat P\equiv -i\partial_{\pi_P} 
\;\;, \end{equation} 
the Hamiltonian and constraint operators, respectively, are given by: 
\begin{eqnarray}\label{Hgen}
\widehat {\cal H}&=&-\pi_Q\cdot\widehat P+\pi_P\cdot V'(\widehat Q) 
\;\;, \\ [1ex] \label{Cgen} 
\widehat C&=&\delta (\frac{1}{2}\widehat P^2+V(\widehat Q)-\epsilon ) 
\;\;, \end{eqnarray}  
where, of course, $V'(Q)\equiv\nabla_QV(Q)$, and the wave function is 
considered as a function $\psi (\pi_P,\pi_Q)$ of the indicated vectors. 

The previous oscillator example suggests to perform a Fourier transformation 
to variables $x,y$, such that the eigenvalue problem becomes: 
\begin{equation}\label{eigengen} 
\widehat {\cal H}\psi (x,y)=\Big (x\cdot (-i\partial_y)-V'(y)\cdot(-i\partial_x)\Big )\psi (x,y) 
= E\psi (x,y) 
\;\;, \end{equation} 
while the constraint operator equation turns into an algebraic constraint: 
\begin{equation}\label{Calgebr}  
\widehat C\psi (x,y)=\delta (\frac{1}{2}x^2+V(y)-\epsilon )\psi (x,y)=0 
\;\;. \end{equation}  
In agreement with the general result (\ref{Hconserv}), we easily 
confirm here that $\widehat {\cal H}\widehat C\psi =\widehat C\widehat {\cal H}\psi$. 
The constraint equation then simply states that the phase space variables are 
constrained to a constant energy surface of the underlying classical system.   
  
The first order 
quasi-linear partial differential equation (\ref{eigengen}) 
can be studied by the method of characteristics \cite{CH}.  
Thus, one finds one equation taking care of the inhomogeneity (right-hand 
side), which can be trivially integrated. Furthermore, the remaining $2n$ 
equations for the characteristics present nothing but the classical Hamiltonian 
equations of motion.   
  
It follows that integrable classical models can (in principle) be decoupled  
in this context of the characteristic equations by canonical transformations.  
This assumes that we can apply them freely at the pre-quantum level, which 
might not be the case. 
It would lead us essentially to the collection of harmonic oscillators 
mentioned at the beginning of Section\,7.1, and corresponding quantum harmonic 
oscillators as studied there. 

Classical crystal-like models with only harmonic forces, or free field theories, respectively, 
will thus give rise to corresponding free quantum mechanical systems here. These are  
constructed in a different way in \cite{tHooft01}. Presumably, the 
(fixing of a large class of)  
gauge transformations invoked there can be related to the existence of integrals of 
motion implied by integrability here. In any case, we conclude that in the present framework 
truly interacting quantum (field) theories might be connected with nonintegrable deterministic systems beneath.
  
Furthermore, we emphasize that the Hamiltonian equations of motion preclude motion into  
classically forbidden regions of the underlying system. Nevertheless, {\it quantum mechanical tunneling}  
is an intrinsic property of the quantum oscillator models that we obtained, as well as of the 
anharmonic oscillator example considered in the following. Similarly, {\it spreading of wave packets} 
is to be expected in the latter case.  
 
In order to demonstrate additional features 
of the eigenvalue problem of (\ref{eigengen}), we concentrate on 
one degree of freedom 
with phase space coordinates $p,q$  
and with a generic anharmonic potential.   

Since the potential depends only on $q$, by locally stretching or squeezing the 
coordinate, i.e. by an ``oscillator transformation'' $q\equiv f(\bar q)$, we can bring 
it into oscillator form, such that $V(f(\bar q))=(1/2)\bar q^2$. Implementing this 
type of transformation, the equations for one degree of freedom are: 
\begin{eqnarray}\label{eigengen1} 
\frac{-i}{f'(\bar q)}\Big (p\partial_{\bar q}-\bar q\partial_p\Big )\psi (p,\bar q)=E\psi (p,\bar q) 
\;\;, \\ [1ex] \label{Calgebr1}
\delta\Big (\frac{1}{2}(p^2+\bar q^2)-\epsilon\Big )\psi (p,\bar q)=0
\;\;, \end{eqnarray} 
with $f'$ denoting the derivative of $f$. This is very much oscillator-like indeed and, 
once more employing polar coordinates, we obtain:  
\begin{eqnarray}\label{eigengen2} 
\frac{-i}{f'(\rho\sin\phi)}\partial_\phi\psi (\rho ,\phi )=E\psi (\rho ,\phi ) 
\;\;, \\ [1ex] \label{Calgebr2}
\delta (\rho^2-\epsilon )\psi (\rho ,\phi )=0
\;\;. \end{eqnarray} 
The eigenvalue problem seems underdetermined. As it stands, it would give rise to an 
unbound continuous spectrum, with no groundstate in particular. 

This apparent defect persists for any number of degrees of freedom.   
However, as we have seen already, sense can be made of the Hamilton operator   
by a suitable regularization, especially by discretizing the phase space 
coordinates. The principles of such 
regularization we still do not know, other than either preserving  
or intentionally breaking symmetries. 

\subsection{An Anharmonic Oscillator}
It is worth while to consider one more example, a onedimensional system with  
Hamiltonian: 
\begin{equation}\label{Hlinpot} 
H\equiv\frac{1}{2}p^2+V_0|q| 
\;\;, \end{equation} 
in order to demonstrate the subtleties associated with regularization.  
For the linear potential, the coordinate dependence of the operator 
on the left-hand side of (\ref{eigengen2}) is 
mild, since $f'(\rho\sin\phi )=V_0^{-1}\rho\sin\phi$, 
and the eigenvalues of its discretized counterpart can be found as follows. 
 
Conveniently discretizing the angular variable as 
$\phi_n\equiv(2\pi n/N)+(3\pi /2)$, $1\leq n\leq N$,     
the eigenvalue equation becomes: 
\begin{equation}\label{lineigen} 
\prod_{n=1}^{N}\Big (1-\lambda\cos (2\pi n/N)\Big )=1\;\;,\;\;\; 
\lambda\equiv 2\pi i\sqrt\epsilon E/NV_0
\;\;. \end{equation} 
Setting $\lambda\equiv 2z/(1+z^2)$, and employing a known identity for the 
finite product arising here, the eigenvalue equation can be transformed into: 
$(1+z^N)^2=(1+z^2)^N$. With hindsight, we choose $N\equiv 2(4N'+1)$ and set 
$z\equiv +\sqrt{u-1}$, 
to obtain: 
\begin{equation}\label{lineigen1} 
u^{N/2}+(1-u)^{N/2}=1 
\;\;. \end{equation}  
This equation has the nice property that, if $u$ is a solution, then so is $1/u$. 
The location of the solutions $u=0,1,\infty$ suggests to look for further solutions  
in the form of $u\equiv\exp 2i\alpha$. Thus, combining the equations for $u$ and $1/u$, 
we arrive at the transcendental equation:
\begin{equation}\label{lineigen2}  
2\sin (\alpha N/2)=(2\sin\alpha )^{N/2} 
\;\;. \end{equation}
From the multitude of its solutions, due to periodicity, we need to find 
$N$ solutions of (\ref{lineigen}). 

Closer inspection shows that, in the limit of large $N$, solutions of (\ref{lineigen2}) 
consist essentially of those zeros of $\sin (\alpha N/2)$ which lie inside the 
intervals $[n\pi -\pi /6,n\pi +\pi /6]$, with integer $n$. Thus, positive energy solutions 
will be obtained momentarily from: 
\begin{equation}\label{esols} 
E_\pm =NV_0\exp (i\frac{\pi}{4}\pm\frac{3}{2}i\alpha ) 
\Big ((2/\epsilon)\sin (\pm\alpha )\Big )^{1/2}
\;\;, \end{equation} 
where either ``$+$'' or ``$-$'' has to be chosen consistently, corresponding to 
the solutions coming in pairs $\exp \pm 2i\alpha$. 

A remark is in order here. Considering only the positive root above, $z\equiv +\sqrt{u-1}$, we avoided 
negative energy solutions. However, there is a price to pay: careful counting reveils that the 
positive energy spectrum is doubly degenerate. The finite positive part is obtained 
from (\ref{esols}), incorporating $\alpha\equiv 2\pi m/N$: 
\begin{eqnarray}\label{esols1}  
&E&=NV_0\exp (3i\pi m/N)\Big ((2/\epsilon )\sin (2\pi m/N)\Big )^{1/2} 
,\;0\leq m_0\leq m\leq N/12 
\;, \\ [1ex] \label{esols2}
&\;&\stackrel{N\rightarrow\infty}{\longrightarrow}\;
\tilde V_0(m+m_0-1)^{1/2} 
\;\;,\;\;m\in\mathbf{N}
\;\;, \end{eqnarray} 
where $m_0$ is an arbitrary constant, within the allowed range, which defines the 
zeropoint energy of the emergent quantum model. The continuum limit is to be taken such 
that $\tilde V_0\equiv V_0(4\pi N/\epsilon )^{1/2}$ stays finite.
The additional solutions can be chosen in a way that their real parts  
move to $+\infty$, as $N\rightarrow\infty$.\footnote{
Corresponding eigenfunctions, i.e. $N$-component discrete eigenvectors, are obtained by 
evaluating products of the kind appearing in (\ref{lineigen}), with $k-1\leq N-1$ 
factors for the $k$th component and a constant for $k=1$.} 

We remark that the spectrum of (\ref{esols2}) differs from the one obtained for the same 
potential in standard quantum mechanics, where WKB yields: $E\propto (m-1/4)^{2/3}$.  

Summarizing, the various illustrated features promise to make 
genuinely interacting models quite difficult to analyze.  
We hope that more interesting results will be obtained with the help of spectrum generating algebras 
or some to-be-developed perturbative methods.      

\section{Conclusions} 
We pursue the view that quantum mechanics is an emergent description of nature, 
which possibly can be based on classical, pre-quantum concepts.  

Our approach is motivated by a construction of a reparametrization-invariant time. 
In turn, this is based on the observation that ``time passes'' when there is an observable change, which is 
localized with the observer. More precisely, necessary are incidents, i.e. observable unit changes, which are 
recorded, and from which invariant quantities characterizing the change of the evolving system can be derived. 

We recall the model of \cite{E03}, invoking compactified extradimensions in which 
a particle moves in addition 
to its relativistic motion in Minkowski space. We 
employ a window to these extradimensions, i.e., we consider a quasi-local 
detector which registers the particle trajectory passing by. Counting  
such incidents, we construct an invariant measure of time.  
   
A basic ingredient is the assumption of ergodicity, such that the system explores 
dynamically the whole allowed energy surface in phase space. This assures that there 
are sufficiently frequent observable incidents. They reflect properties  
of the dynamics with respect to (subsets of) Poincar\'e sections. Roughly, the  
passing time corresponds to the observable change there. Then, the particle's proper time 
is linearly related to the physical time, however, subject to stochastic fluctuations.            

Thus, the reparametrization-invariant time    
based on quasi-local observables naturally induces stochastic  
features in the behavior of the external relativistic particle motion. Due to  
quasi-periodicity (or, generally, more strongly irregular features) of the emerging discrete time,  
the remaining predictable aspects appear as in unitary  
quantum mechanical evolution.  

In reparametrization-invariant, ``timeless'' single-particle systems, this idea has been realized in various 
forms \cite{ES02,E03}. Presently, this has led us to assume the relation between the constructed 
physical time $t$ and standard proper time $\tau$ of the evolving system in the form of a statistical distribution,     
$P(\tau ;t)=P(\tau -t)$, cf. (\ref{P}). Here we assume that the distribution is not 
explicitly time-dependent, which means, the physical clock is decoupled from the system under study. 
We explore the consequences of this situation for the description of the system. 
  
We have shown how to introduce ``states'', eventually building up a Hilbert space, in terms of certain 
functional integrals, (\ref{state})--(\ref{adjstate}), which arise from the study of a suitable 
classical generating functional. The latter was introduced earlier in a 
different context, studying classical mechanics in functional form \cite{GRT,GR00}. We employ this  
as a convenient tool, and modify it, in order to describe the observables of reparametrization-invariant systems 
with discrete time (Section\,6). 
Studying the evolution of the states in general (Section\,5), we are led to the Schr\"odinger equation,  
(\ref{Schroedinger}). However, the Hamilton operator (\ref{Hamiltonian}) has a non-standard first-order 
form with respect to phase space coordinates.   
  
The choice of boundary conditions of the classical paths contributing to the generating 
functional plays a crucial role and deserves better understanding. 

Furthermore, illustrating the emergent 
quantum models in various examples, we demonstrate that proper regularization of the continuum Hamilton  
operator is indispensable, in order that well-defined quantum mechanical systems emerge, with bounded spectra 
and a stable groundstate, in particular. Most desirable is a deeper undertanding of this mapping between 
the continuum Hamilton operator, which is straightforward to write down, given a classical pre-quantum system, 
and the effective quantum mechanics, which emerges after proper regularization only. Especially, limitations 
imposed by symmetries and consistency of the procedure need further study. 

It is a common experience that 
the preservation of continuum symmetries through discretization is difficult, for example, see 
\cite{Snyder,TDL83,JN97,Pullinetal}. We wonder, whether other regularization schemes are conceivable. 
The possiblity, mentioned after (\ref{Schroedinger}), that the need for regularization is an artefact 
of decoupled clock degrees of freedom deserves further study.    
  
We find that truly interacting quantum (field) theories might be connnected to 
nonintegrable classical models beneath, since otherwise the degrees of freedom represented in  
the stationary Schr\"odinger equation here, can principally be decoupled by employing classical 
canonical transformations. 

Finally, we come back to the probabilistic relation between physical time and the evolution parameter 
figuring in the parameterized classical equations of motion, which is the underlying raison d'\^etre of the  
presented stroboscopic quantization. One would like to include the clock degrees of freedom consistently 
into the dynamics, in order to address the closed Universe. This can be achieved 
by introducing suitable projectors into the generating functional. Their task is to replace a  
simple quasi-local detector which responds to a particle trajectory passing 
through in Yes/No fashion; by counting such incidents, an invariant measure of time 
has been obtained before \cite{ES02,E03}. In a more general setting, this detector/projector has to 
be defined in terms of observables of the closed system. In this way, typical conditional probabilities 
can be handled, such as describing ``What is the probability of observable $X$ having a value 
in a range $x$ to $x+\delta x$, {\it when} observable $Y$ has value $y$?''. Criteria for selecting the 
to-be-clock degrees of freedom are still unknown, other than simplicity. Most   
likely the resulting description of evolution and implicit notion of physical time will correspond to 
our distribution $P(\tau ;t)$ of (\ref{P}), however, now evolving explicitly with the system.   
We leave this for future study.       

The stroboscopic quantization emerging from underlying classical dynamics 
may be questioned in many respects. It might violate one or the other assumption of existing 
no-go theorems relating to hidden variables theories. However, we believe it is 
interesting to learn more about working examples, before discussing this. Unitary evolution,  
tunneling effects, and spreading of wave packets are recovered in this framework.  
Interacting theories remain to be explored.


    %

    \end{document}